# Phonon Transport in Graphene

## Denis L. Nika[+] and Alexander A. Balandin[*]

**Department of Electrical Engineering and Materials Science and Engineering Program,**

**Bourns College of Engineering, University of California, Riverside, CA 92521 U.S.A.**

### INVITED REVIEW

### *Abstract*

Properties of phonons – quanta of the crystal lattice vibrations – in graphene have attracted strong attention of the physics and engineering communities. Acoustic phonons are the main heat carriers in graphene near room temperature while optical phonons are used for counting the number of atomic planes in Raman experiments with few-layer graphene. It was shown both theoretically and experimentally that transport properties of phonons, i.e. energy dispersion and scattering rates, are substantially different in the quasi two-dimensional system such as graphene compared to basal planes in graphite or three-dimensional bulk crystals. The unique nature of two-dimensional phonon transport translates to unusual heat conduction in graphene and related materials. In this review we outline different theoretical approaches developed for phonon transport in graphene, discuss contributions of the in-plane and cross-plane phonon modes and provide comparison with available experimental thermal conductivity data. Particular attention is given to analysis of recent theoretical results for the phonon thermal conductivity of graphene and few-layer graphene, and the effects of the strain, defects and isotopes on the phonon transport in these systems.

---

[+]On leave from the Department of Theoretical Physics, State University of Moldova, Republic of Moldova

[*]Corresponding author email address (A.A.B.): balandin@ee.ucr.edu

## I. Introduction

Heat removal became a crucial issue for continuing progress in electronic industry owing to increased levels of dissipated power density and speed of electronic circuits [1]. Self-heating is a major problem in optoelectronics and photonics. These facts stimulated recent interest to thermal properties of materials. Acoustic phonons – quanta of the crystal lattice vibrations – are the main heat carriers in a variety of material systems. The phonon and thermal properties of nanostructures are substantially different from those of bulk crystals [2-14]. Semiconductor nanowires do not conduct heat as well as bulk crystals due to increased phonon - boundary scattering [3-4] or changes in the phonon dispersion and density of states (DOS) [2, 5-7]. The thermal conductivity $K$ of thin films and nanowires is usually lower than that of the corresponding bulk materials owing to the extrinsic effects such as phonon – boundary scattering [4, 8-9]. However, theoretical studies suggested that phonon transport in strictly two-dimensional (2D) and one-dimensional (1D) systems can revealed exotic behavior, leading to infinitely large intrinsic thermal conductivity [10-11]. These theoretical results led to discussions of the validity of Fourier's law in low-dimensional systems [15-16] and further stimulated interest to the acoustic phonon transport in 2D systems.

In this review we focus on the specifics of the acoustic phonon transport in graphene. After a brief summary of the basics of thermal physics in nanostructures and experimental data for graphene's thermal conductivity, we discuss, in more details, various theoretical approaches to the phonon thermal conductivity in graphene. A special attention is given to the analysis of the most recent theoretical and computational results on relative contributions of different phonon polarization branches to the thermal conductivity in graphene. The readers more interested in the experimental thermal conductivity values of graphene and related materials in the general context of carbon allotropes are referred to a different review [17].

## II. Basics of Phonon Transport and Thermal Conductivity

The main experimental technique for investigation of the acoustic phonon transport in a given material system is measurement of its lattice thermal conductivity. In this section, we define the main characteristics of heat conduction. The thermal conductivity is introduced through Fourier's law:

$$\vec{\phi} = -K \nabla T \,, \tag{1}$$

where $\vec{\phi}$ is the heat flux, $\nabla T$ is the temperature gradient and $K = (K_{ij})$ is the thermal conductivity tensor. In the isotropic medium the thermal conductivity does not depend on the direction of the heat flow and $K$ is treated as a constant. The latter is valid for the small temperature variations only. In a wide temperature range, the thermal conductivity is a function of temperature, i.e. $K \equiv K(T)$. In general, in solid materials heat is carried by phonons and electrons so that $K = K_p + K_e$, where $K_p$ and $K_e$ are the phonon and electron contributions, respectively. In metals or degenerately-doped semiconductors, $K_e$ is dominant due to large concentration of free carriers. The value of $K_e$ can be determined from the measurement of the electrical conductivity $\sigma$ via the Wiedemann-Franz law:

$$\frac{K_e}{\sigma T} = \frac{\pi^2 k_B^2}{3e^2} \,, \tag{2}$$

where $k_B$ is the Boltzmann's constant and $e$ is the charge of an electron. The phonons are usually the main heat carriers in carbon materials. Even in graphite, which has metal-like properties [18], the heat conduction is dominated by acoustic phonons [19]. This fact is explained by the strong covalent sp$^2$ bonding resulting in high in-plane phonon group velocities and low crystal lattice unharmonicity for in-plane vibrations.

The phonon thermal conductivity can be written as

$$K_p = \Sigma_j \int C_j(\omega) \upsilon_j^2(\omega) \tau_j(\omega) d\omega \,, \tag{3}$$

where summation is performed over the phonon polarization branches $j$, which include two transverse acoustic and one longitudinal acoustic branches, $\upsilon_j = d\omega_j / dq$ is the phonon group velocity of the $j$th branch, which, in many solids, can be approximate by the sound velocity, $\tau_j$ is the phonon relaxation time, $C_j = \hbar \omega_j \partial N_0 (\hbar \omega_j / k_B T) / \partial T$ is the contribution to heat capacity from the $j$th branch and $N_0(\frac{\hbar \omega_j}{k_B T}) = [\exp(\frac{\hbar \omega_j}{k_B T}) - 1]^{-1}$ is the Bose-Einstein phonon equilibrium distribution function. The phonon mean-free path (MFP) $\varLambda$ is related to the relaxation time through the expression $\varLambda = \tau \upsilon$. In the relaxation-time approximation (RTA), various scattering mechanisms, which limit MFP, are additive, i.e. $\tau_j^{-1} = \sum_i \tau_{i,j}^{-1}$, where $i$ denotes scattering mechanisms. In typical solids, the acoustic phonons, which carry bulk of heat, are scattered by other phonons, lattice defects, impurities, conduction electrons and interfaces [20-23].

In the ideal crystals, i.e. crystals without lattice defects or rough boundaries, $\varLambda$ is limited by the phonon - phonons scattering due to the crystal lattice anharmonicity. In this case, the thermal conductivity is referred to as intrinsic. The anharmonic phonon interactions, which lead to the finite thermal conductivity in three dimensions, can be described by the Umklapp processes [20]. The Umklapp scattering rates depend on the Gruneisen parameter $\gamma$, which determines degree of the lattice anharmonicity [20-21]. Thermal conductivity is extrinsic when it is mostly limited by the extrinsic effects such phonon – rough boundary or phonon – defect scattering.

In nanostructures the phonon energy spectra are quantized due to spatial confinement of the acoustic phonons. The quantization of phonon energy spectra usually leads to decreasing phonon group velocity. The modification of the phonon energies, group velocities and density of states together with phonon scattering from boundaries affect the thermal conductivity of nanostructures. In most of cases, the spatial confinement of acoustic phonons results in the reduction of the phonon thermal conductivity [24-25]. However, it was predicted that the thermal conductivity of nanostructures embedded within the acoustically hard barrier layers can be increased via spatial confinement of acoustic phonons [6, 9, 26].

The phonon boundary scattering can be evaluated as [23]

$$\frac{1}{\tau_{B,j}} = \frac{\upsilon_j}{D} \frac{1-p}{1+p}, \tag{4}$$

where $D$ is the nanostructure or grain size and $p$ is the specularity parameter defined as a probability of specular scattering at the boundary. The momentum-conserving specular scattering ($p$=1) does not add to thermal resistance. Only diffuse phonon scattering from rough interfaces ($p \rightarrow 0$), which changes the momentum, limits the phonon MFP. One can find $p$ from the surface roughness or use it as a fitting parameter to experimental data. The commonly used expression for the phonon specularity was given by Ziman [23]

$$p(\lambda) = \exp(-\frac{16\pi^3\eta^2}{\lambda^2}), \tag{5}$$

where $\eta$ is the root mean square deviation of the height of the surface from the reference plane and $\lambda$ is the length of the incident phonon wave.

In the case when the phonon - boundary scattering is dominant, thermal conductivity scales with the nanostructure or grain size $D$ as $K_p \sim C_p \upsilon \Lambda \sim C_p \upsilon^2 \tau_B \sim C_p \upsilon D$. In the very small structures with $D << \Lambda$, the thermal conductivity dependence on the physical size of the structure becomes more complicated due to the strong quantization of phonon energy spectra [24]. The specific heat $C_p$ depends on the phonon density of states, which leads to different $C_p(T)$ dependence in three-dimensional (3D), two-dimensional and one-dimensional systems, and reflected in $K(T)$ dependence at low $T$ [20, 23]. In bulk at low $T$, $K(T) \sim T^3$ while it is $K(T) \sim T^2$ in 2D systems.

The thermal conductivity $K$ defines how well a given material conducts heat. Another characteristics – the thermal diffusivity $\alpha$ – defines how fast the material conducts heat. The thermal diffusivity is given by the expression:

$$\alpha = \frac{K}{C_p \rho_m}, \tag{6}$$

where $\rho_m$ is the mass density. Many experimental techniques measure the thermal diffusivity rather than thermal conductivity.

### III.   Experimental Data for Thermal Conductivity of Graphene

We start by providing a brief summary of experimental data available for the thermal conductivity of graphene. The first measurements of heat conduction in graphene [27-32] were carried out at UC Riverside in 2007 (see figure 1). The investigation of the phonon transport was made possible with development of the optothermal Raman measurement technique. The experiments were performed with the large-area suspended graphene layers exfoliated from the high-quality Kish and highly ordered pyrolytic graphite. It was found that the thermal conductivity varies in a wide range and can exceed that of the bulk graphite, which is ~2000 W/mK at room temperature (RT). It was also determined that the electronic contribution to heat conduction in the un-gated graphene near RT is much smaller than that of phonons, i.e. $K_e \ll K_p$. The phonon in graphene MFP was estimated to be on the order of 800 nm near RT [28].

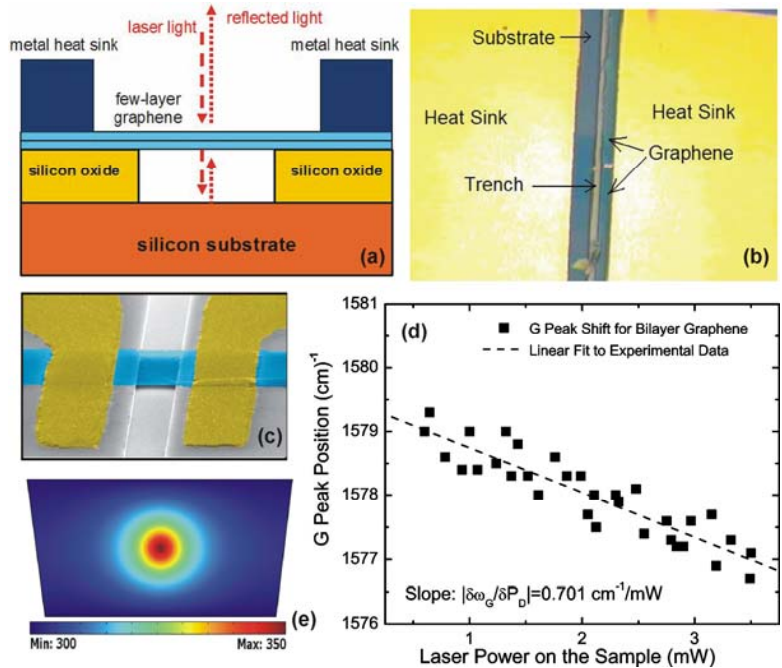

**Figure 1**:  (a) Schematic of the thermal conductivity measurement showing suspended FLG flakes and excitation laser light. (b) Optical microscopy images of FLG attached to metal heat sinks. (c) Colored scanning electron microscopy image of the suspended graphene flake to clarify a typical structure geometry. (d) Experimental data for Raman G-peak position as a function of laser power, which determines the local temperature rise in response to the dissipated power. (e) Finite-element simulation of temperature distribution in the flake with the given geometry used to extract the thermal conductivity. Figure is after Ref. [30] reproduced with permission from the Nature Publishing Group.

Several independent studies, which followed, also utilized the Raman optothermal technique but modified it via addition of a power meter under the suspended portion of graphene. It was found that the thermal conductivity of suspended high-quality chemical vapour deposited (CVD) graphene exceeded ~2500 W/mK at 350 K, and it was as high as $K \approx 1400$ W/mK at 500 K [33]. The reported value was also larger than the thermal conductivity of bulk graphite at RT. Another Raman optothermal study with the suspended graphene found the thermal conductivity in the range from ~1500 to ~5000 W/mK [34]. Another group that repeated the Raman-based measurements found $K \approx 630$ W/mK for the suspended graphene membrane [35]. The differences in the actual temperature of graphene under laser heating, strain distribution in the suspended graphene of various sizes and geometries can explain the data variation.

Another experimental study reported the thermal conductivity of graphene to be ~1800 W/mK at 325 K and ~710 W/mK at 500 K [36]. These values are lower than that of bulk graphite. However, instead of measuring the light absorption in graphene under conditions of their experiment, the authors of Ref. [36] assumed that the optical absorption coefficient should be 2.3%. It is known that due to many-body effects, the absorption in graphene is the function of wavelength $\lambda$, when $\lambda > 1$ eV [37-39]. The absorption of 2.3% is observed only in the near-infrared at ~1 eV. The absorption steadily increases with decreasing $\lambda$ (increasing energy). The 514.5-nm and 488-nm Raman laser lines correspond to 2.41 eV and 2.54 eV, respectively. At 2.41 eV the absorption is about $1.5 \times 2.3\% \approx 3.45\%$ [38]. The value of 3.45% is in agreement with the one reported in another independent study [40]. Replacing the assumed 2.3% with 3.45% in the study reported in Ref. [36] gives ~2700 W/mK at 325 K and 1065 W/mK near 500 K. These values are higher than those for the bulk graphite and consistent with the data reported by other groups [33, 40], where the measurements were conducted by the same Raman optothermal technique but with the measured light absorption.

The data for suspended or partially suspended graphene is closer to the intrinsic thermal conductivity because suspension reduces thermal coupling to the substrate and scattering on the substrate defects and impurities. The thermal conductivity of fully supported graphene is smaller. The measurements for exfoliated graphene on $SiO_2$/Si revealed in-plane $K \approx 600$ W/mK near RT [41]. Solving Boltzmann's transport equation (BTE) and comparing with their experiments, the authors determined that the thermal conductivity of free graphene should be ~3000 W/mK near RT.

Despite the noted data scatter in the reported experimental values of the thermal conductivity of graphene, one can conclude that it is very large compared to that for bulk silicon ($K$=145 W/mK at RT) or bulk copper ($K$=400 W/mK at RT) – important materials for electronic applications. The differences in $K$ of graphene can be attributed to variations in the graphene sample lateral sizes (length and width), thickness non-uniformity, material quality (e.g. defect concentration and surface contaminations), grain size and orientation, as well as strain distributions. Often the reported thermal conductivity values of graphene corresponded to different sample temperatures $T$, despite the fact that the measurements were conducted at ambient. The strong heating of the samples was required due to the limited spectral resolution of the Raman spectrometers used for temperature measurements. Naturally, the thermal conductivity values determined at ambient but for the samples heated to $T$~350 K and $T$~600 K over substantial portion of their area would be different and cannot be directly compared. One should also note that the data scatter for thermal conductivity of carbon nanotubes (CNTs) is much larger than that for graphene. For a more detail analysis of the experimental uncertainties the readers are referred to a comprehensive review [17].

## IV.   Phonon Scattering and Transport in Suspended Few-Layer Graphene

The phonon thermal conductivity undergoes an interesting evolution when the system dimensionality changes from 2D to 3D. This evolution can be studied with the help of suspended few-layer graphene (FLG) with increasing thickness $H$ – number of atomic planes $n$. It was reported in Ref. [30] that thermal conductivity of suspended uncapped FLG decreases with increasing $n$ approaching the bulk graphite limit (see figure 2). This trend was explained by considering the intrinsic quasi-2D crystal properties described by the phonon Umklapp scattering [30]. As $n$ in FLG increases – the phonon dispersion changes and more phase-space states become available for phonon scattering leading to thermal conductivity decrease. The phonon scattering from the top and bottom boundaries in suspended FLG is limited if constant $n$ is maintained over the layer length. The small thickness of FLG ($n$<4) also means that phonons do not have transverse component in their group velocity leading to even weaker boundary scattering term for the phonons. In thicker FLG films the boundary scattering can increase due to the non-zero cross-plane phonon velocity component. It is also harder to maintain the constant thickness through the whole area of FLG flake. These factors can lead to the thermal conductivity below the graphite limit. The graphite value recovers for thicker films.

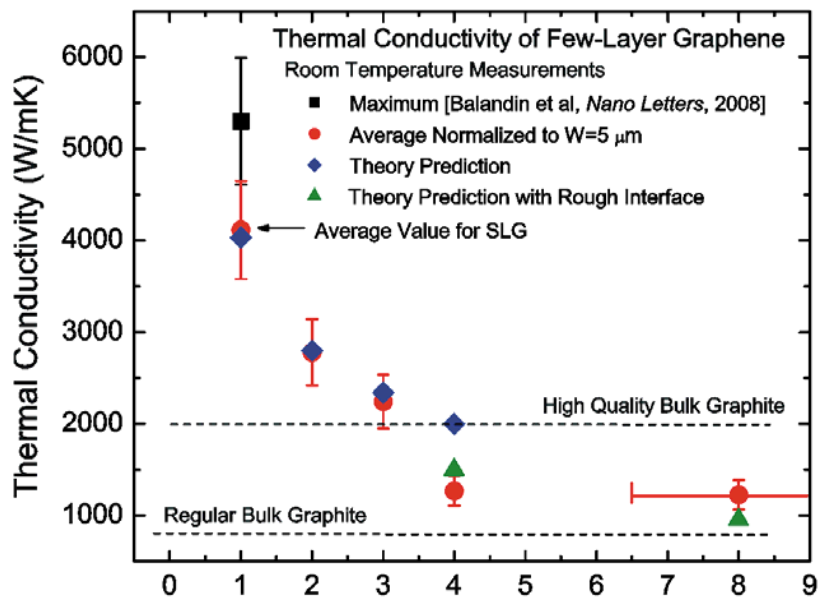

**Figure 2:** Measured thermal conductivity as a function of the number of atomic planes in FLG. The dashed straight lines indicate the range of bulk graphite thermal conductivities. The blue diamonds were obtained from the first-principles theory of thermal conduction in FLG based on the actual phonon dispersion and accounting for all allowed three-phonon Umklapp scattering channels. The green triangles are Callaway–Klemens model calculations, which include extrinsic effects characteristic for thicker films. Figure is after Ref. [30] reproduced with permission from the Nature Publishing Group.

The experimentally observed evolution of the thermal conductivity in FLG with $n$ varying from 1 to $n\sim4$ [30] is in agreement with the theory for the crystal lattices described by the Fermi-Pasta-Ulam Hamiltonians [42]. The molecular-dynamics (MD) calculations for graphene nanoribbons with the number of planes $n$ from 1 to 8 [17] also gave the thickness dependence of the thermal conductivity in agreement with the UC Riverside experiments [30]. The strong reduction of the thermal conductivity as $n$ changes from 1 to 2 is in line with the earlier theoretical predictions [44]. In another reported study, the Boltzmann's transport equation was solved under the assumptions that in-plane interactions are described by Tersoff potential while Lennard-Jones potential models interactions between atoms belonging to different layers [45-46]. The obtained results suggested a strong thermal conductivity decrease as $n$ changed from 1 to 2 and slower decrease for $n>2$.

The thermal conductivity dependence on the FLG is entirely different for the encased FLG where thermal transport is limited by the acoustic phonon scattering from the top and bottom boundaries and disorder. The latter is common when FLG is embedded between two layers of dielectrics. An experimental study [43] found $K\approx160$ W/mK for encased single-layer graphene (SLG) at $T\approx310$ K. It increases to ~1000 W/mK for graphite films with the thickness of 8 nm. It was also found that the suppression of thermal conductivity in encased graphene, as compared to bulk graphite, was stronger at low temperature where $K$ was proportional to $T^{\beta}$ with $1.5<\beta<2$ [43]. Thermal conduction in encased FLG was limited by the rough boundary scattering and disorder penetration through graphene.

## V.    Phonon Spectra in Graphene, FLG and Graphene Nanoribbons

Intriguing thermal and electrical properties of graphene, FLG [17, 27–30, 47-48] and graphene nanoribbons (GNRs) [49-51] stimulates investigations of phonon energy spectra in these materials and structures [52-66]. The phonon energy spectrum is important for determining the sound velocity, phonon density of states, phonon-phonon or electron-phonon scattering rates, lattice heat capacity, as well as the phonon thermal conductivity. The optical phonon properties manifest themselves in Raman measurements. The number of graphene layers, their quality and stacking order can be clearly distinguished using the Raman spectroscopy [30, 67-70]. For these reasons, significant efforts have been made to accurately determine the phonon energy dispersion in graphite [52-55], graphene [45, 56-61, 66], GNRs [62-65, 71], and to reveal specific features of their phonon modes.

The phonon dispersion in graphite along $\Gamma - M - K - \Gamma$ directions (see figure 3(a), where the graphene Brillouin zone is shown) measured by X-ray inelastic scattering was reported in Refs. [52-53]. A number of research groups calculated the phonon energy dispersion in graphite, graphene and GNRs using various theoretical approaches, including continuum model [64-65], Perdew-Burke-Ernzerhof generalized gradient approximation (GGA) [52, 54-55], first-order local density function approximation (LDA) [54, 56, 60], fourth- and fifth-nearest neighbor force constant (4NNFC and 5NNFC) approaches [53, 55, 61], Born-von

Karman or valence force field (VFF) model of the lattice dynamics [57-58, 66], utilized the Tersoff and Brenner potentials [59] or Tersoff and Lennard-Jones potentials [45-46]. All these models are based on different sets of the fitting parameters, which are determined from comparison with the experimental phonon dispersion [52-53,72].

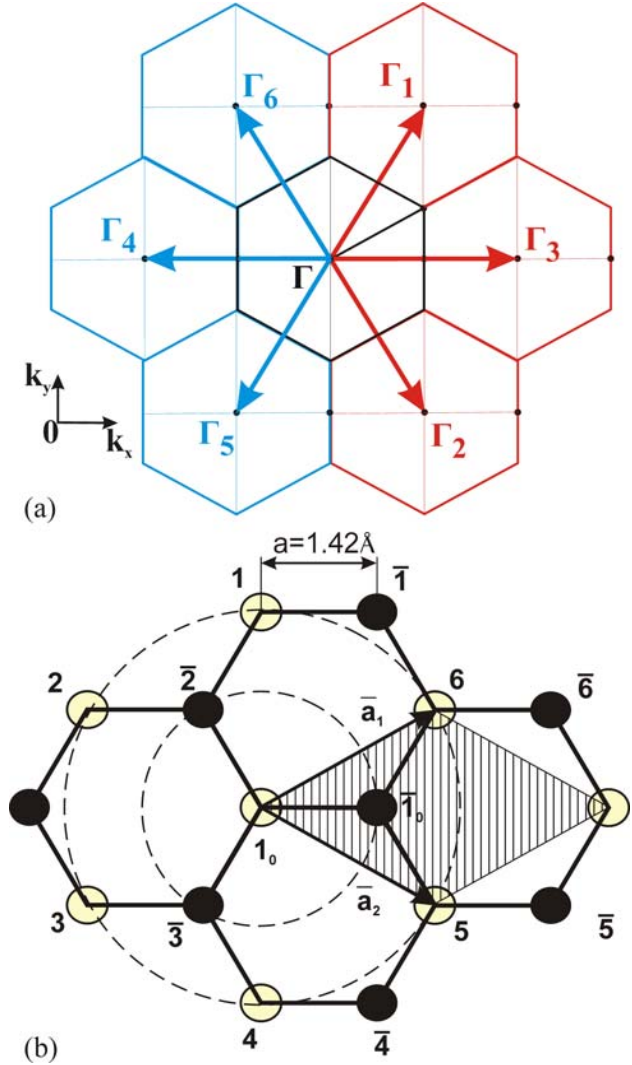

**Figure 3:** (a) Reciprocal lattice of graphene. (b) Graphene crystal lattice. The rhombic unit cell is shown as a shaded region. Figure is reproduced from Ref. [66] with permission from the American Physical Society.

The number of parameters in the theoretical models depends on the model specifics and the number of the considered atomic neighbors. The number of the parameters varies from 5 [55] to 23 [61]. For example, our VFF model for graphene used only six parameters [66]. In this model, all interatomic forces are resolved into bond-streching and bond-bending forces [73-75]. This model takes into account stretching and bending interactions with two in-plane and two out-of-plane atomic neighbors as well as doubled streching-streching interactions with the nearest in-plane neighbors. The honeycomb crystal lattice of graphene utilized in this model is presented in figure 3(b). The rhombic unit cell of graphene, shown as a dashed region, contains two atoms and is defined by two basis vectors $\vec{a}_1 = a(3,\sqrt{3})/2$, and $\vec{a}_2 = a(3,-\sqrt{3})/2$, where $a = 0.142$ nm is the distance between two nearest carbon atoms. The six phonon polarization branches $s = 1,\ldots, 6$ in SLG are shown in figure 4. These branches are (i) out-of-plane acoustic (ZA) and out-of-plane optical (ZO) phonons with the displacement vector along the Z axis; (ii) transverse acoustic (TA) and transverse optical (TO) phonons, which corresponds to the transverse vibrations within the graphene plane; (iii) longitudinal acoustic (LA) and longitudinal optical (LO) phonons, which corresponds to the longitudinal vibrations within the graphene plane.

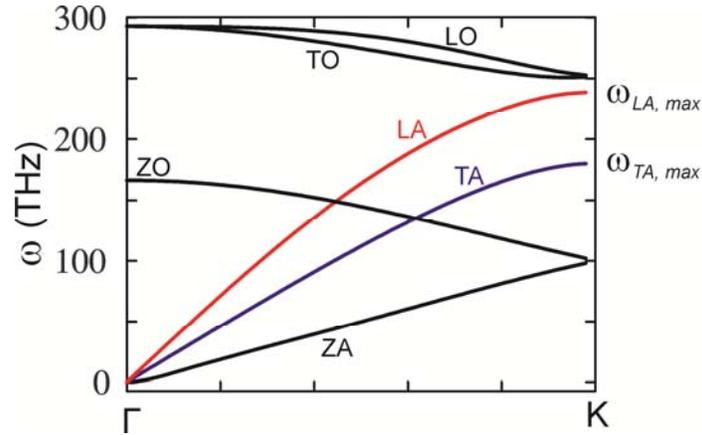

**Figure 4:** Phonon frequencies $\omega_s$ in graphene calculated using the valence force field model. Figure is reproduced from Ref. [29] with permission from the Institute of Physics and Deutsche Physikalische Gesellschaft.

Although various theoretical models are in a qualitative agreement with each other, they predict substantially different phonon frequencies in $\Gamma$, M or K points of the Brillouin zone. Moreover, some of the models give the same frequencies for the LO - LA phonons [55-56,59] and ZO - TA phonons [53-54,57,66] at M point while the rest of the models predict non-equal frequencies for these phonons at M point [52,58,60]. The comparison between phonon frequencies at the high-symmetry points of Brillouin zone is presented in Tables I and II. The discrepancy in the calculated phonon dispersion can easily translate to the differences in the predicted thermal conductivity values. Specifically, the relative contribution of the LA, TA and ZA phonons to heat transport can be varied in a wide range depending on specifics of the phonon dispersion used.

The unit cell of the $n$-layer graphene contains $2 \cdot n$ atoms, therefore $6 \cdot n$ quantized phonon branches appear in $n$-layer graphene. In figure 5(a-b) we show the phonon dispersions in bilayer graphene. Weak van der Waals interaction between monolayers leads to the coupling of long wavelength phonons only and quantization of the low-energy part of the spectrum with $q<0.1q_{max}$ for LA, TA, LO, TO and ZO phonons (see figure 5(b)) and with $q<0.4q_{max}$ for ZA phonons. The modification of phonon energy spectrum in $n$-layer graphene as compared with that in single layer graphene results in substantial change of three phonon scatterings and reduction of the intrinsic thermal conductivity in $n$-layer graphene.

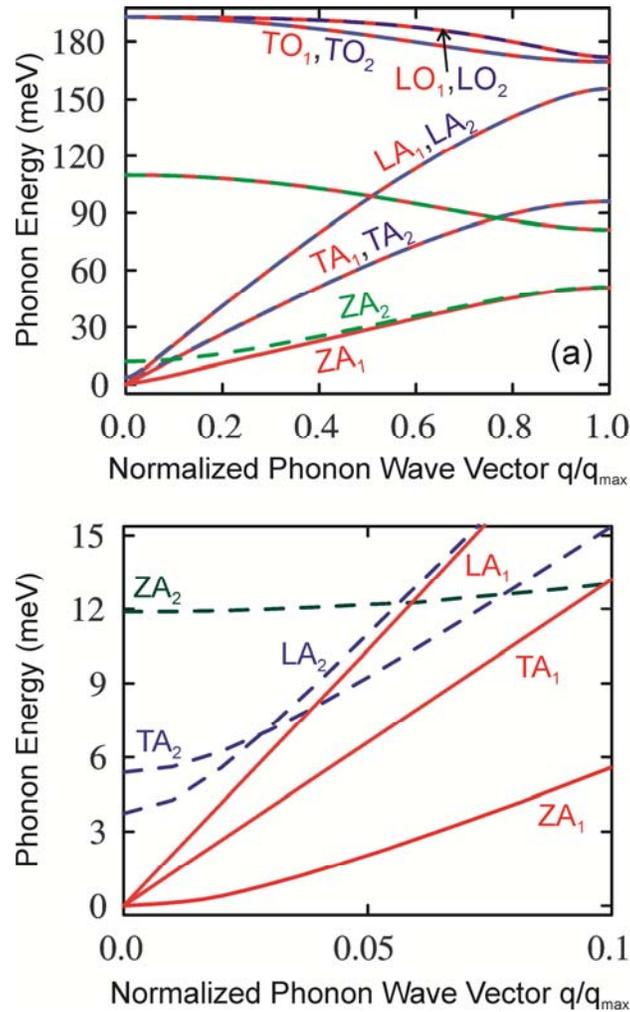

**Figure 5:** Phonon energy spectra in bilayer graphene calculated using the valence force field model shown for (a) $\Gamma - M$ direction and (b) near the Brillouin zone center. Figure is after Ref. [30] reproduced with permission from the Nature Publishing Group.

## VI.   Acoustic Phonon Transport in Two-Dimensional Crystals

We now address in more details some specifics of the acoustic phonon transport in 2D systems. Investigation of heat conduction in graphene [27-28] and CNTs [76] raised the issue of ambiguity in definition of the intrinsic thermal conductivity for 2D and 1D crystal lattices. It was theoretically shown that the intrinsic thermal conductivity limited by the crystal anharmonisity has the finite value in 3D bulk crystals [11, 42]. However, many theoretical models predict that the intrinsic thermal conductivity reveals a logarithmic divergence in strictly 2D systems, $K \sim ln(N)$, and the power-law divergence in 1D systems, $K \sim N^{\alpha}$, with the number of atoms $N$ ($0 < \alpha < 1$) [11, 15, 42, 76-80]. The logarithmic divergence can be removed by introduction of the *extrinsic* scattering mechanisms such as scattering on defects or coupling to the substrate [42]. Alternatively, one can define the *intrinsic* thermal conductivity of a 2D crystal for a given size of the crystal.

Graphene is not an ideal 2D crystal, considered in most of the theoretical works, since graphene atoms vibrate in three directions. Nevertheless the intrinsic graphene thermal conductivity strongly depends on the graphene sheet size due to weak scatterings of low-energy phonons by other phonons in the system, which consists only of one atomic plane. Therefore, the phonon boundary scattering is an important mechanism for phonon relaxation in graphene. The most recent studies suggested that an accurate accounting of the higher-order anharmonic processes, i.e. above three-phonon Umklapp scattering, and inclusion of the normal phonon processes into consideration allow one to limit the low-energy phonon MFP. The normal phonon processes do not contribute directly to thermal resistance but affect the phonon mode distribution [81-82]. However, even these studies found that the graphene sample has to be very large (>10 μm) to obtain the size-independent thermal conductivity.

The specific phonon transport in the quasi - 2D system such as graphene can be illustrated with an expression derived by Klemens specifically for graphene [19, 83]. In the framework of BTE approach and RTA, the intrinsic Umklapp-limited thermal conductivity of graphene can be written as [19, 83]:

$$K = \frac{\rho_m}{2\pi\gamma^2} \frac{\bar{\upsilon}^4}{f_m T} \ln(\frac{f_m}{f_B}). \tag{7}$$

Here $f_m$ is the upper limit of the phonon frequencies defined by the phonon dispersion, $\bar{\upsilon}$ is the average phonon group velocity, $f_B = \left( M\bar{\upsilon}^3 f_m / 4\pi\gamma^2 k_B TL \right)^{1/2}$ is the size-dependent low-bound cut-off frequency for acoustic phonons, introduced by limiting the phonon MFP with the graphene layer size $L$.

In Ref. [84] we improved equation (7) by taking into account the actual maximum phonon frequencies and Gruneisen parameters $\gamma_s$ ($s=TA$, $LA$) determined separately for LA and TA phonon branches. The Gruneisen parameters were computed by averaging the phonon mode-dependent $\gamma_s(\bar{q})$ for all relevant phonons (here $\bar{q}$ is the wave vector):

$$K = \frac{1}{4\pi k_B T^2 h} \sum_{s=TA,LA} \int_{q_{\min}}^{q_{\max}} \{[\hbar\omega_s(q)\frac{d\omega_s(q)}{dq}]^2 \tau_{U,s}^K(q) \frac{exp[\hbar\omega_s(q)/k_B T]}{[exp[\hbar\omega_s(q)/k_B T]-1]^2} q\}dq \, . \qquad (8)$$

Here $\hbar\omega_s(q)$ is the phonon energy, $h = 0.335$ nm is the graphene layer thickness and $\tau_{U,s}^K(q)$ is the three-phonon mode-dependent Umklapp relaxation time, which was derived using an expression from Refs. [19-20] but introducing separate life-times for $LA$ and $TA$ phonons:

$$\tau_{U,s}^K = \frac{1}{\gamma_s^2} \frac{M\bar{\upsilon}_s^2}{k_B T} \frac{\omega_{s,\max}}{\omega^2} \, , \qquad (9)$$

where $s=TA$, $LA$, $\bar{\upsilon}_s$ is the average phonon velocity for a given branch, $\omega_{s,\max} = \omega(q_{\max})$ is the maximum cut-off frequency for a given branch and $M$ is the mass of an atom. In Refs. [19, 83-84] the contribution of ZA phonons to thermal transport has been neglected because of their low group velocity and large Gruneisen parameter $\gamma_{ZA}$ [54, 84]. Equation (9) can be used to calculate the thermal conductivity with the actual dependence of the phonon frequency $\omega_s(q)$ and the phonon velocity $d\omega_s(q)/dq$ on the phonon wave number. To simplify the model one can use the liner dispersion $\omega_s(q)=\bar{\upsilon}_s q$ and re-write it as:

$$K_U = \frac{\hbar^2}{4\pi k_B T^2 h} \sum_{s=TA,LA} \int_{\omega_{\min}}^{\omega_{\max}} \{\omega^3 \tau_{U,s}^K(\omega) \frac{exp[\hbar\omega/kT]}{[exp[\hbar\omega/kT]-1]^2}\}d\omega \, . \qquad (10)$$

Substituting equation (9) to equation (10) and performing integration one obtains

$$K_U = \frac{M}{4\pi T h} \sum_{s=TA,LA} \frac{\omega_{s,\max} \overline{\upsilon}_s^2}{\gamma_s^2} F(\omega_{s,\min}, \omega_{s,\max}), \tag{11}$$

where

$$F(\omega_{s,\min}, \omega_{s,\max}) = \int_{\hbar\omega_{s,\min}/k_BT}^{\hbar\omega_{s,\max}/k_BT} \xi \frac{exp(\xi)}{[exp(\xi)-1]^2} d\xi =$$

$$[ln\{exp(\xi)-1\} + \frac{\xi}{1-exp(\xi)} - \xi]\Big|_{\hbar\omega_{s,\min}/k_BT}^{\hbar\omega_{s,\max}/k_BT} \tag{12}$$

In the above equation, $\xi = \hbar\omega / k_BT$, and the upper cut-off frequencies $\omega_{s,\max}$ are defined from the actual phonon dispersion in graphene (see figure 4): $\omega_{LA,\max} = 2\pi f_{LA,\max}(\Gamma M) = 241$ THz, $\omega_{TA,\max} = 2\pi f_{TA,\max}(\Gamma M) = 180$ THz.

The integrand in equation (12) can be further simplified near RT when $\hbar\omega_{s,\max} > k_BT$, and it can be expressed as

$$F(\omega_{s,\min}) \approx -ln\{|\, exp(\hbar\omega_{s,\min}/k_BT)-1\,|\} + \frac{\hbar\omega_{s,\min}}{k_BT} \frac{exp(\hbar\omega_{s,\min}/k_BT)}{exp(\hbar\omega_{s,\min}/k_BT)-1}. \tag{13}$$

There is a clear difference in the heat transport in basal planes of bulk graphite and in single layer graphene [19, 83]. In the former the heat transport is approximately two-dimensional only till some low-bound cut-off frequency $\omega_{\min}$. Below $\omega_{\min}$ there appears strong coupling with the cross-plane phonon modes and heat starts to propagate in all directions, which reduces the contributions of these low-energy modes to heat transport along basal planes to negligible values. In bulk graphite there is a physically reasonable reference point for the on-set of the cross-plane coupling, which is the ZO' phonon branch near ~4 THz observed in the spectrum of bulk graphite. The presence of ZO' branch and corresponding $\omega_{\min} = \omega_{ZO'}(q = 0)$ allows one to avoid the logarithmic divergence in the Umklapp-limited thermal conductivity integral (see equations (10–13)) and calculate it without considering other scattering mechanisms.

The physics of heat conduction is principally different in graphene where the phonon transport is 2D all the way to zero phonon frequency $\omega(q=0)=0$. There is no on-set of the cross-plane heat transport at the long-wavelength limit in the system, which consists of only one atomic plane. This is no $ZO'$ branch in the phonon dispersion of graphene (see figure 4). Therefore the low-bound cut-off frequencies $\omega_{s,\min}$ for each $s$ are determined from the condition that the phonon MFP cannot exceed the physical size $L$ of the flake, i.e.

$$\omega_{s,\min} = \frac{\bar{\upsilon}_s}{\gamma_s} \sqrt{\frac{M\bar{\upsilon}_s}{k_B T} \frac{\omega_{s,\max}}{L}} \ . \tag{14}$$

We would like to emphasize here that using size-independent graphite $\omega_{\min}$ for SLG or FLG (as has been proposed in Ref. [85]) is without of scientific merit and leads to the erroneous calculation of thermal conductivity [86]. Equations (12-14) constitute a simple analytical model for calculation of the thermal conductivity of graphene layer, which retains such important features of graphene phonon spectra as different $\bar{\upsilon}_s$ and $\gamma_s$ for $LA$ and $TA$ branches. The model also reflects the two-dimensional nature of heat transport in graphene all the way down to zero phonon frequency.

In figure 6 we present the dependence of thermal conductivity of graphene on the dimension of the flake $L$. The data is presented for the averaged values of Gruneisen parametera $\gamma_{LA}$=1.8 and $\gamma_{TA}$=0.75 obtained from *ab initio* calculations, as well as for several other close sets of $\gamma_{LA,TA}$ to illustrate the sensitivity of the result to Gruneisen parameters. For small graphene flakes, $K$ dependence on $L$ is rather strong. It weakens for flakes with $L \geq 10$ μm. The calculated values are in good agreement with available experimental data for suspended exfoliated [27-28] and CVD graphene [33-34]. The horizontal line indicates the experimental thermal conductivity for bulk graphite, which is exceeded by graphene's thermal conductivity at smaller $L$. Thermal conductivity, presented in figure 6, is an *intrinsic* quantity limited by the three-phonon Umklapp scattering only. But it is determined for a specific graphene flake size since $L$ defines the low-bound (long-wavelength) cut-off frequency in Umklapp scattering through equation (14). In experiments, thermal conductivity is also limited by defect scattering. When the size of the flake becomes very large with many polycrystalline grains, the scattering on their boundaries will also lead to phonon relaxation. The latter can be included in our model through adjustment of $L$. The extrinsic phonon scattering mechanisms or high-order phonon-phonon scatterings prevent indefinite growth of thermal conductivity of graphene with $L$.

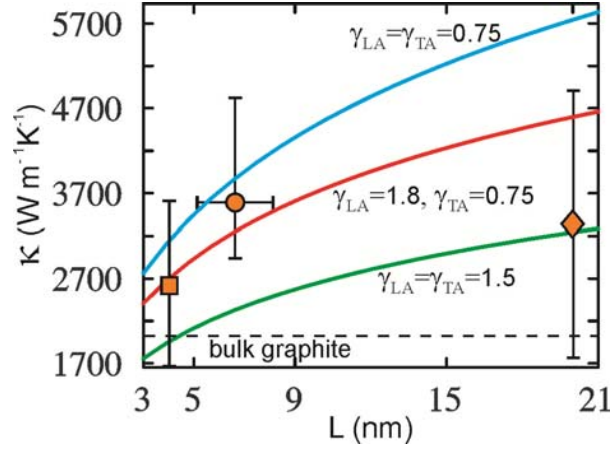

**Figure 6:** Calculated room temperature thermal conductivity of graphene as a function of the lateral size for several values of Gruneisen parameter. An experimental data points from Refs. [27-28] (circle), [33] (square) and [34] (rhomb) are shown for comparison.

## VII. Q-Space Diagram Theory of Phonon Transport in Graphene

The simple models described in the previous section are based on the Klemens-like expressions for the relaxation time (see equation (9)). Therefore they do not take into account all peculiarities of the 2D three-phonon Umklapp processes in SLG or FLG, which are important for the accurate description of the thermal transport. There are two types of the three-phonon Umklapp scattering processes [20]. The first type is the scattering when a phonon with the wave vector $\vec{q}(\omega)$ absorbs another phonon from the heat flux with the wave vector $\vec{q}'(\omega')$, i.e. the phonon leaves the state $\vec{q}$. For this type of scattering processes the momentum and energy conservation laws are written as:

$$\vec{q} + \vec{q}' = \vec{b}_i + \vec{q}'' , \; i = 1,2,3$$
$$\omega + \omega' = \omega''$$

(15)

The processes of the second type are those when the phonons $\vec{q}$ of the heat flux decay into two phonons with the wave vectors $\vec{q}'$ and $\vec{q}''$ leaving the state $\vec{q}$, or, alternatively, two phonons $\vec{q}'(\omega')$ and $\vec{q}''(\omega'')$ merge together forming a phonon with the wave vector $\vec{q}(\omega)$, which correspond to the phonon coming to the state $\vec{q}(\omega)$. The conservation laws for this type are given by:

$$\vec{q} + \vec{b}_i = \vec{q}' + \vec{q}'', \quad i = 4,5,6$$
$$\omega = \omega' + \omega'', \tag{16}$$

In equations (15-16) $\vec{b}_i = \Gamma\vec{\Gamma}_i$, $i = 1,2,...,6$ is one of the vectors of reciprocal lattice (see figure 3(a)).

Calculations of the thermal conductivity in graphene taking into account all possible three-phonon Umklapp processes allowed by the equations (15-16) and actual phonon dispersions were carried out for the first time in Ref. [66]. For each phonon mode ($q_i$, $s$), were found all pairs of the phonon modes ($\vec{q}',s'$) and ($\vec{q}'',s''$) such that the conditions of equations (15-16) are met. As a result, in ($\vec{q}'$)-space were constructed the *phase diagrams* for all allowed three-phonon transitions [66]. Using the long-wave approximation (LWA) for a matrix element of the three-phonon interaction authors of Ref. [66] obtained for the Umklapp scattering rates

$$\frac{1}{\tau_U^{(I),(II)}(s,\vec{q})} = \frac{\hbar\gamma_s^2(\vec{q})}{3\pi\rho\upsilon_s^2(\vec{q})} \sum_{s's'';b_i} \iint \omega_s(\vec{q})\omega_{s'}'(\vec{q}')\omega_{s''}''(\vec{q}'') \times$$
$$\{N_0[\omega_{s'}'(\vec{q}')] \mp N_0[\omega_{s''}''(\vec{q}'')]\} + \frac{1}{2} \mp \frac{1}{2}\} \times \delta[\omega_s(\vec{q}) \pm \omega_{s'}'(\vec{q}') - \omega_{s''}''(\vec{q}'')]dq_l'dq_\perp'. \tag{17}$$

Here $q_l'$ and $q_\perp'$ are the components of the vector $\vec{q}'$ parallel or perpendicular to the lines defined by equations (15-16), correspondingly, $\gamma_s(\vec{q})$ is the mode-dependent Gruneisen parameter, which is determined for each phonon wave vector and polarization branch and $\rho$ is the surface mass density. In equation (17) the upper signs correspond to the processes of the first type while the lower signs correspond to those of the second type. The integrals for $q_l, q_\perp$ are taken along and perpendicular to the curve segments, correspondingly, where the conditions of equations (15-16) are met. Integrating along $q_\perp$ in equation (17) one can obtain the line integral

$$\frac{1}{\tau_U^{(I),(II)}(s,\vec{q})} = \frac{\hbar\gamma_s^2(\vec{q})\omega_s(\vec{q})}{3\pi\rho\upsilon_s^2(\vec{q})} \sum_{s',s'',b} \int_l \frac{\pm(\omega_{s'}''-\omega_s)\omega_{s''}''}{\upsilon_{\perp,s'}(\omega_{s'}')}(N_0' \mp N_0'' + \frac{1}{2} \mp \frac{1}{2})dq_l'. \qquad (18)$$

The phonon scattering on the rough edges of graphene can be evaluated using equation (4). The total phonon relaxation rate is given by:

$$\frac{1}{\tau_{tot}(s,q)} = \frac{1}{\tau_U(s,q)} + \frac{1}{\tau_B(s,q)}. \qquad (19)$$

The sensitivity of the room temperature thermal conductivity, calculated using equations (17-19), to the value of the specular parameter of phonon boundary scattering is illustrated in figure 7. The data is presented for different sizes (widths) of the graphene flakes. The experimental data points for suspended exfoliated [27-28] and CVD [33-34] graphene are also shown for comparison. Table III provides representative experimental and theoretical data for the suspended and supported graphene.

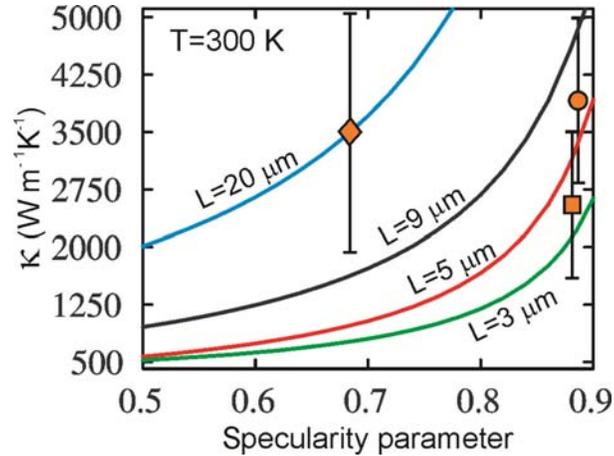

**Figure 7:** Calculated room temperature thermal conductivity of the suspended graphene as a function of the specularity parameter $p$ for the phonon scattering from the flake edges. Note a strong dependence on the size of the graphene flakes. An experimental data points from Refs. [27-28] (circle), [33] (square) and [34] (rhomb) are shown for comparison.

**VIII. Phonon Transport and Thermal Conductivity of Graphene Nanoribbons**

Measurements of thermal properties of graphene stimulated a surge of interest to theoretical and experimental studies of heat conduction in graphene nanoribbons [49-51, 71, 87-95]. It is important to understand how lateral sizes affect the phonon transport properties from both fundamental science and practical applications point of view. In the last few years a number of theoretical works investigated phonon transport and heat conduction in graphene nanoribbons with various lengths, widths, edge roughness and defect concentrations. The authors used MD simulations [49-51, 87-90], nonequilibrium Green's function method [91-93] and BTE approaches [71, 94].

Keblinsky and co-workers [49] found from the MD study that the thermal conductivity of graphene is $K \approx 8000 - 10000$ W/mK at RT for the square graphene sheet. The $K$ value was size independent for $L>5$ nm [49]. For the ribbons with fixed $L=10$ nm and width $W$ varying from 1 to 10 nm, $K$ increased from ~1000 W/mK to 7000 W/mK. The thermal conductivity in GNR with rough edges can be suppressed by orders of magnitude as compared to that in GNR with perfect edges [49, 51]. The isotopic superlattice modulation of GNR or defects of crystal lattices also significantly decreases the thermal conductivity [92-93]. The uniaxial stretching applied in the longitudinal direction enhances the low-temperature thermal conductance for the 5 nm arm-chair or zigzag GNR up to 36 % due to the stretching-induced convergence of phonon spectra to the low-frequency region [91].

Aksamija and Knezevic [71] calculated the dependence of the thermal conductivity of GNR with the width 5 nm and RMS edge roughness $\Delta = 1$ nm on temperature (see figure 8). The thermal conductivity was calculated taking into account the three-phonon Umklapp, mass-defect and rough edge scatterings [71]. The authors obtained RT thermal conductivity $K \sim 5500$ W/mK for the graphene nanoribbon. The study of the nonlinear thermal transport in rectangular and triangular GNRs under the large temperature biases was reported in Ref. [95]. The authors found that in short (~6 nm) rectangular GNRs, the negative differential thermal conductance exists in a certain range of the applied temperature difference. As the length of the rectangular GNR increases the effect weakens. A computational study reported in Ref. [96] predicted that the combined effects of the edge roughness and local defects play a dominant role in determining the thermal transport properties of zigzag GNRs. The experimental data on thermal transport in GNR is very limited. In Ref. [97] the authors used an electrical self-heating methods and extracted the thermal conductivity of sub 20-nm GNRs to be more than 1000 W/mK at 700 – 800 K. The calculated and measured data for the thermal conductivity of graphene nanoribbons is also given in Table III.

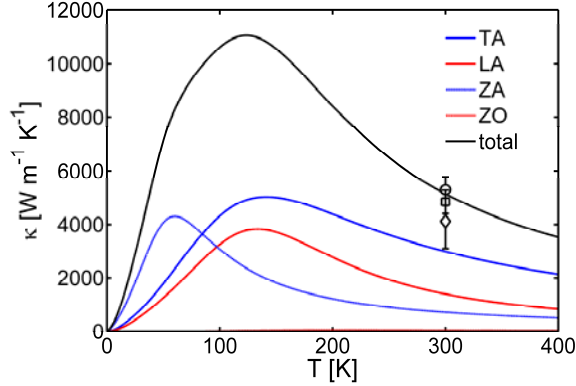

**Figure 8:** Thermal conductivity results for GNRs of width W=5 nm and rms edge roughness Δ=1 nm, showing contributions from individual phonon branches TA, LA, ZA, and ZO and total. An experimental data points from Refs. [27-28] are shown for comparison. Figure is reproduced from Ref. [71] with permission from the American Institute of Physics.

## IX.  Analysis of Recent Theoretical and Computational Results

In this section we review and analyze the most recent theoretical results pertinent to phonon transport in graphene. Ong and Pop [98] examined thermal transport in graphene supported on $SiO_2$ using MD simulations. The approach employed by the authors utilized the reactive empirical bond order (REBO) potential to model the atomic interaction between the C atoms, Munetoh potential to model the atomic interactions between the Si and O atoms and Lennard-Jones potential to model the van der Waals type C-Si and C-O couplings. Surprisingly, Ong and Pop [98] found from their calculations that increasing the strength of the graphene-substrate interaction enhances the heat flow and effective thermal conductivity along the supported graphene. The authors attributed this result to the coupling of graphene ZA modes to the substrate Rayleigh waves, which linearizes the phonon dispersion, increases the group velocity of the hybridized modes and, thus, enhances the thermal flux. This conclusion is an opposite of the results reported by Seol et al. [41] where they found that the coupling to the substrate leads to additional phonon scattering with the corresponding reduction of the in-plane thermal conductivity of graphene.

Qiu and Ruan [99-100] addressed the problem of relative contributions of ZA phonons to thermal transport in the framework of the equilibrium MD simulations. Their conclusion was that in suspended SLG out-of-plane ZA phonons are coupled with in-plane phonons due to the 3$^{rd}$-order and higher-order anharmonic interactions, which results in about $25 - 30\%$ contribution of ZA phonons to the thermal conductivity of graphene. In supported SLG the contribution of all acoustic and ZO phonon branches are reduced owing to the SLG-substrate interface scattering and breakdown of the symmetries for both in-plane and out-of-plane phonons. The contributions of ZA phonons to thermal conductivity are suppressed stronger than the contributions of TA and LA phonons. Qiu and Ruan [99-100] stated that the in-plane TA and LA phonons are the dominant heat carriers in supported SLG and make major contribution in suspended SLG.

The theory of the lattice thermal conductivity in graphene and few-layer graphene based on a numerical solution of the linearized BTE for phonons was proposed by Singh et al. [46]. The authors used the force fields described by the optimized Tersoff interatomic potential for the in-plane interactions and Lennard-Jones potential for the out-of plane interactions. It was found within this model's approximations that the out-of-plane ZA phonons make dominant contributions to the thermal conductivity of SLG and FLG. The thermal conductivity reduction with the increasing number of graphene layers was explained by reinforcement of ZA phonons scattering due to breaking of the SLG selection rules and corresponding suppression of ZA phonon contribution to the thermal conductivity. The Singh et al. [46] conclusions are in contrast to the Qiu and Ruan's [99-100] results obtained from MD simulations. One should note here that the fast reduction of the thermal conductivity in FLG with the increasing number of atomic planes, experimentally observed by Ghosh et al. [30], has been explained by various theoretical approaches with substantially different set of assumptions [30, 45-46, 101-102].

The strain effects on the thermal conductivity of graphene and GNRs were studied computationally in Ref. [103]. Authors used MD simulations and found that the thermal conductivity of graphene is very sensitive to the tensile strain. It was shown that when the strains are applied in both directions – parallel and perpendicular – to the heat transfer path, the graphene sheets undergo complex reconstructions. As a result, some of the strained graphene structures can have higher thermal conductivity than that of SLG without strain [103]. The suggested strong strain dependence of the thermal conductivity of graphene can explain some of discrepancies in the reported experimental values of the thermal conductivity of suspended graphene. The suspended graphene flakes and membranes are expected to have different strain field depending on the size and geometry of the suspected graphene sample.

The strong dependence of the thermal conductivity of graphene on the defect concentration was established in the computational studies reported in Refs. [104-105]. Both studies used MD simulations. According to Hao et al. [104] 2 % of the vacancies or other defects can reduce the thermal conductivity of graphene by as much as a factor of five to ten. Zhang et al. [105] determined from their MD simulations that the thermal conductivity of pristine graphene should be ~2903 W/mK at RT. According to their calculations the thermal conductivity of graphene can be reduced by a factor of 1000 at the vacancy defect concentration of ~9 %. The numeric results of Refs. [104-105] suggest another possible explanation of the experimental data scatter, which is different defect density in the examined graphene samples. For example, if the measurements of the thermal conductivity of graphene by the thermal bridge technique give smaller values than those by the Raman optothermal technique, one should take into account that the thermal bridge technique requires substantial number of fabrication steps, which result in residual defects.

An intriguing question in the theory of phonon transport in graphene is a relative contribution to heat conduction by LA, TA and ZA phonon polarization branches. The calculations of the thermal conductivity from BTE within RTA [19, 83-84, 106] or by using the three-phonon matrix elements obtained from the LWA [66] show relatively small (down to negligible) contributions of ZA phonons. The latter is mainly attributed to the large (negative) Gruneisen parameter and the small phonon group velocity. The calculations performed in the framework of the linearized BTE and three-phonon matrix elements based on the third-order interatomic force constants (IFCs) claim that the heat conduction is dominated by the ZA phonons [45-46, 107]. This conclusion results from the mode-dependent third-order IFCs and the selection rule in ideal graphene, which restricts the phase space for the phonon-phonon scattering, thus, increasing the ZA modes lifetime. However, graphene placement on any substrate and presence of nanoscale corrugations in graphene lattice can break the symmetry selection rule, which restricts ZA phonon scattering. It is also possible that ZA dispersion undergoes modification, e.g. linearization, due to the substrate coupling as discussed in Ref. [98].

Singh et al. [82] provided critical review of existing theoretical approaches and concluded that the Klemens-type approximations to scattering matrix elements fail for graphene because these approaches do not include special selection rules, which restrict anharmoning scatterings of ZA and ZO phonons. The authors emphasized the importance of the three-phonon normal processes for accurate description of the thermal conductivity of graphene and suggested that heat conduction in graphene is dominated by the out-of-plane ZA phonons. The theoretical approach of Singh et al. [82] is based on the three-phonon matrix elements derived without assumptions of LWA. This approach involved the third-order anharmonic IFCs written as

$$\Phi_{\alpha\beta\gamma}(0k,l'k',l''k'') = \frac{\partial^3 V}{\partial u_\alpha(0k)\partial u_\beta(l'k')\partial u_\gamma(l''k'')},$$ (20)

where $u_\alpha(lk)$ is the $\alpha$th component of a displacement of the $k$th atom in the $l$th unit cell and $V = V(c_1, c_2, ... c_N)$ is the empirical interatomic potential. The number of the constants in a set $(c_1, c_2, ... c_N)$ depends on the type of the interatomic potential, while values of the constants are usually determined from the comparison of theoretically calculated cohesive energy, phonon energy or another measurable quantity with experimental one. For the calculation of $\Phi_{\alpha\beta\gamma}(0k,l'k',l''k'')$ Singh et al. [82] employed the Tersoff interatomic potential with the set of parameters determined from the best fit of the theoretical phonon energies to the available experimental data for graphite (see Refs. [46, 59]). Nevertheless the optimized Tersoff potential give a poor agreement with the available experimental frequencies for ZO phonons near Γ point (difference is about 80 rad/ps at Γ point), ZO, LO and TA phonons near M point (difference is about 40-50 rad/ps for both phonons at M point) and ZA, TA and LO phonons near K point (difference is about 20 rad/ps for ZA, 40 rad/ps for TA and 100 rad/ps for LO phonons at K point) (see figure 1 from Ref. [59]). Moreover, the phonon energies and group velocities of TA phonons, calculated using this potential, are overestimated over a half of the Brilluoin zone. In Ref. [82] the phonon energies were found as a solution of a set of equations of motions which depends on the second-order (harmonic) IFCs only (see equation (1) from Ref. [46])

$$\Phi_{\alpha\beta}(0k,l'k') = \frac{\partial^2 V}{\partial u_\alpha(0k)\partial u_\beta(l'k')}.$$  (21)

Therefore the calculation of the third-order IFCs $\Phi_{\alpha\beta\gamma}(0k,l'k',l''k'')$, which are important for the thermal transport and for determining the relative contribution of LA, TA, ZA phonons to the thermal conductivity, is not a well justified procedure and can lead to wrong conclusions. It is known that the elastic and vibration properties depend strongly on the type of the empirical potential as shown in many theoretical publications [108-110]. Broido et al. [108] demonstrated that Tersoff and the environmental-dependent interatomic potentials give vastly different thermal expansion and Gruneisen coefficients. Cowley [109] analyzed vibration properties of silicon using Stillinger-Weber and Tersoff potentials and concluded that none of these potentials provide a fully satisfactory description of the lattice vibrations. Sevencli et al. [110] demonstrated that LA and TA modes in the hybrid boron nitride – graphene sheets are equally well described by the Tersoff potential and the fourth-nearest-neighbor forces constants while the energies of ZA, ZO, TO and LO phonons are not. The higher-order phonon processes can also change the relative contribution of different phonons to the thermal conductivity. For example, Qiu and Ruan [99-100] predicted strong coupling of ZA phonons with LA and TA phonons due to the higher-order Umklapp and normal processes with the corresponding increase of their scattering. In their calculation ZA phonons accounted for 15% in the graphene on the substrate and ~25-30% in the suspended SLG [99-100].

For the reasons discussed above, we consider the question of the relative contributions of different phonon polarizations to the thermal conductivity of graphene to be still open. Experimentally, it is difficult to address this question. The measurements of temperature dependence of the thermal conductivity cannot present evidence in favor of one or the other phonon contribution because $K(T)$ dependence in graphite is known to be strongly influenced by the material quality [43, 111-112].

## X. Isotope Effects on Phonon Spectra and Transport in Graphene

Naturally occurring carbon materials are made up of two stable isotopes of $^{12}$C (~99%) and $^{13}$C (~1%). The change in isotope composition modifies dynamic properties of crystal lattices and affects their thermal conductivity. The isotopically purified materials are characterized by enhanced thermal conductivity [113-117]. The knowledge of isotope effects on thermal properties is valuable for understanding the phonon transport. The isotope composition affects directly the phonon relaxation rates in the phonon mass-difference scattering processes. The phonon-scattering rate on point defects, $1/\tau_P$, is given as [19, 106] $1/\tau_P \propto V_0(\omega^\alpha / \upsilon_j^\beta)\Upsilon$, where $V_0$ is the volume per one atom in the crystal lattice, $\Upsilon$ is the strength of the phonon - point defect scattering, $\alpha$=3(4) and $\beta$=2(3) for 2D (3D) system, correspondingly. In the perturbation theory $\Upsilon$ can be written as [19,106]

$$\Upsilon = \sum_i f_i \left[ \left(1 - M_i / \overline{M}\right)^2 + \varepsilon \left( \gamma \left(1 - R_i / \overline{R}\right)\right)^2 \right], \qquad (22)$$

where $f_i$ is the fractional concentration of the substitutional foreign atoms, e.g. impurity, defect or isotope atoms, $M_i$ is the mass of the $i$th substitutional atom, $\overline{M} = \sum_i f_i M_i$ is the average atomic mass, $R_i$ is the Pauling ionic radius of the $i$th foreign atom, $\overline{R} = \sum_i f_i R_i$ is the average radius and $\varepsilon$ is a phenomenological parameter. The mass of a foreign atom – impurity, vacancy, defect or isotope – is well known, while the local displacement $\Delta R = \overline{R} - R_i$ due to the atom radius or bond-length difference is usually not known well.

One can see from equation (22) that the phonon-isotope scattering is unique in a sense that unlike impurity or defect scattering it involves only the well-defined mass-difference term, $\Delta M = \overline{M} - M_i$, without the ambiguous volume or bond-strength difference term, $\Delta R = \overline{R} - R_i$ and $\varepsilon$. As the system dimensionality changes from 3D to 2D, the phonon scattering on point defects undergoes additional modification owing to the different phonon DOS. The change in the phonon DOS reveals itself via dependence of $1/\tau_P$ on $\omega$ and $\upsilon$. Thus, the isotope effects in graphene are particularly important for understanding its thermal properties and, more generally, for development of theory of the phonon transport in low-dimensional systems.

The first experimental study of the isotope effects on the thermal properties of graphene was reported just recently [118]. The isotopically modified graphene containing various percentages of $^{13}$C were synthesized by CVD technique [119-120]. The regions of different isotopic composition were parts of the same graphene sheet to ensure uniformity in material parameters. The thermal conductivity of the isotopically pure $^{12}$C (0.01% $^{13}$C) graphene, determined by the optothermal Raman technique [17, 27-28, 30, 33, 40], was higher than 4000 W/mK at the temperature $T\sim320$ K, and more than a factor of two higher than the value of $K$ in a graphene sheets composed of a 50%-50% mixture of $^{12}$C and $^{13}$C.

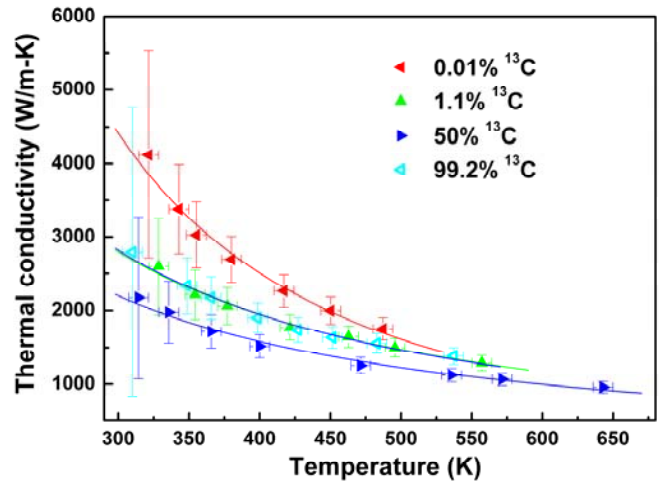

**Figure 9:** Thermal conductivity of the suspended graphene film with $^{13}$C isotope concentrations of 0.01%, 1.1% (natural abundance), 50% and 99.2%, respectively, as a function of the temperature measured with the micro-Raman spectrometer. Note strong dependence of the thermal conductivity on the isotope concentration. Figure is reproduced from Ref. [118] with permission from the Nature Publishing Group.

Figure 9 shows thermal conductivity in the isotopically modified graphene as a function of temperature. The evolution of thermal conductivity with the isotope content was attributed to the changes in the phonon – point defect scattering rate $1/\tau_P$ via the mass-difference term $\Delta M = \overline{M} - M_i$. The phonon $\upsilon$ and mass density do not undergo substantial modification with the isotope composition. The relative change in the phonon velocity $\upsilon_{^{12}C}/\upsilon_{natural}$ is related to the mass densities of the respective lattices $\upsilon_{^{12}C}/\upsilon_{natural} = (M_{natural}/M_{^{12}C})^{1/2}$. Removal of 1% $^{13}$C in natural diamond causes the velocity to increase only by a tiny fraction, which cannot account for the observed strong change in the thermal conductivity.

The reported experimental data in Ref. [118] agrees well with the authors' own MD simulations, corrected for the long-wavelength phonon contributions via the Klemens model [118] and other numeric data reported previously [66, 121-122]. A recent study [123] reported an analytical model and MD simulations of the isotope effects in carbon materials, including nanoribbons. The results of the calculations for the thermal conductivity dependence on the isotope composition are in line with the measurements [118]. It was also predicted theoretically that further reduction in thermal conductivity of the isotopically engineered graphene [124] could be achieved if the isotopes were organized in small size clusters rather than being distributed randomly [125]. These findings are inline with those obtained for rectangular GNRs [126].

## XI.   Conclusions

We reviewed available theoretical and experimental data on phonon transport in graphene. Phonons are the dominant heat carriers in the ungated graphene samples near room temperature. The unique nature of quasi-2D phonons translates to unusual heat conduction in graphene and related materials. Recent computational studies suggest that the thermal conductivity of graphene depends strongly on the concentration of defects and strain distribution. Investigation of the physics of quasi-2D phonons in graphene can shed light on the thermal energy transfer in low-dimensional systems.

*Acknowledgements*

This work was supported, in part, by the National Science Foundation (NSF) projects US EECS-1128304, EECS-1124733 and EECS-1102074, by the US Office of Naval Research (ONR) through award N00014-10-1-0224, Semiconductor Research Corporation (SRC) and Defense Advanced Research Project Agency (DARPA) through FCRP Center on Functional Engineered Nano Architectonics (FENA), and DARPA-DMEA under agreement H94003-10-2-1003. DLN acknowledges the financial support through the Moldova State Project No. 11.817.05.10F.

**Table I: Energies of ZO and LO Phonons at Γ Point in Graphite and Graphene**

| Sample | $\Gamma_{ZO}$ (cm$^{-1}$) | $\Gamma_{LO}$ (cm$^{-1}$) | Comments | Refs |
|---|---|---|---|---|
| graphite | --- | 1583[a] | experiment: X-ray scattering | [a]52 [b]53 [c]54 [d]72 [e]55 [f]56 [g]60 |
| graphite | --- | 1581[b] | experiment: X-ray scattering | |
| graphite | 899[c] | 1593[c] | theory: LDA | |
| graphite | ~820[a], 879[c], 881[c] | 1559[c], 1561[c], 1581-1582[a] | theory: GGA | |
| graphite | 868[b] | 1577[b] | theory: 5NNFC | |
| graphite | ~920[d] | ~1610[d] | theory: six-parameter force constant model | |
| graphene | 879[c], 881[c], 884[e] | 1554[c], 1559[c], 1569[e] | theory: GGA | |
| graphene | 890[g], 896[g], ~900[f] | 1586[f], 1595[g], 1597[g] | theory: LDA | |
| graphene | 893 | 1581 | theory: Born-von Karman | 57 |
| graphene | 889[h], 883.5[i] | 1588[h], 1555[i] | theory: VFF model | [h]58 [i]66 |
| graphene | ~1300 | ~1685 | theory: optimized Tersoff | 59 |
| | ~1165 | ~1765 | theory: optimized Brenner | |

**Table II. Phonon Energies at K and M Points in Graphite and Graphene**

| Sample | $K_{ZA}$ (cm$^{-1}$) | $K_{TA}$ (cm$^{-1}$) | $K_{LA}$ (cm$^{-1}$) | Comments | Refs |
|---|---|---|---|---|---|
| graphite | --- | --- | 1194[a] | experiment: X-ray $\omega_{LO}(M) > \omega_{LA}(M)$ | [a]52 |
| graphite | 542[b] | 1007[b] | 1218[b] | experiment: X-ray $\omega_{LO}(M) > \omega_{LA}(M)$; $\omega_{ZO}(M) \approx \omega_{TA}(M)$ | [b]53 |
| graphite | --- | --- | --- | experiment: HREELS $\omega_{LO}(M) > \omega_{LA}(M)$; $\omega_{ZO}(M) < \omega_{TA}(M)$ | 72 |
| graphite | 540[c] | 1009[c] | 1239[c] | theory: LDA; $\omega_{LO}(M) > \omega_{LA}(M)$; $\omega_{ZO}(M) \approx \omega_{TA}(M)$ | |
| graphite | 534[c], 540[c] | ~960[a], 998[c], 999[c] | 1220[a], 1216[c], 1218[c] | theory: GGA; $^{a,c}\omega_{LO}(M) > \omega_{LA}(M)$; $^{c}\omega_{ZO}(M) \approx \omega_{TA}(M)$ | |
| graphite | 542[b] | 1007[b] | 1218[b] | theory: 5NNFC; $\omega_{LO}(M) > \omega_{LA}(M)$; $\omega_{ZO}(M) \approx \omega_{TA}(M)$ | [c]54 [d]55 |
| graphene | 535[c], 539[d] | 997[c], 1004[d] | 1213[c], 1221[d] | theory: GGA; $^{c}\omega_{LO}(M) > \omega_{LA}(M)$; $^{d}\omega_{LO}(M) \approx \omega_{LA}(M)$; $^{c,d}\omega_{ZO}(M) \approx \omega_{TA}(M)$ | |
| graphene | ~520[e,f] | ~990[f] ~1000[e] | ~1250[f], ~1220[e] | theory: LDA; $^{e}\omega_{LO}(M) \approx \omega_{LA}(M)$; $^{e}\omega_{ZO}(M) \approx$ $\approx \omega_{ZA}(M) << \omega_{TA}(M)$; $^{f}\omega_{LO}(M) > \omega_{LA}(M)$; $^{f}\omega_{ZO}(M) > \omega_{ZA}(M)$; | [e]56 [f]60 |
| graphene | 495 | 1028 | 1199 | theory: BvK model $\omega_{LO}(M) > \omega_{LA}(M)$; $\omega_{ZO}(M) \approx \omega_{TA}(M)$ | 57 |
| graphene | 544[g], 532[h] | 1110[g], 957[h] | 1177[g], 1267[h] | theory: VFF model; $^{g,h}\omega_{LO}(M) > \omega_{LA}(M)$; $^{g}\omega_{ZO}(M) < \omega_{TA}(M)$; $^{h}\omega_{ZO}(M) \approx \omega_{TA}(M)$ | [g]58 [h]66 |
| graphene | ~635 | ~1170 | ~1170 | theory: Tersoff $\omega_{LO}(M) \approx \omega_{LA}(M)$; $\omega_{ZO}(M) > \omega_{TA}(M)$ | 59 |
| graphene | ~585 | ~1010 | ~1240 | theory: Brenner $\omega_{LO}(M) > \omega_{LA}(M)$; $\omega_{ZO}(M) > \omega_{TA}(M)$ | |

**Table III: Thermal conductivity of graphene and graphene nanoribons**

| Sample | K (W/mK) | Method | Comments | Refs |
|---|---|---|---|---|
| graphene | ~2000 – 5000 | Raman optothermal | suspended; exfoliated | 27,28 |
| FLG | 1300 - 2800 | Raman optothermal | suspended; exfoliated; n=2-4 | 30 |
| graphene | ~2500 | Raman optothermal | suspended; CVD | 33 |
| graphene | ~1500-5000 | Raman optothermal | suspended; CVD | 34 |
| graphene | 600 | Raman optothermal | suspended; exfoliated; T ~ 660 K | 35 |
| graphene | 600 | Electrical | supported; exfoliated; | 41 |
| FLG nanoribbon | 1100 | Electrical self-heating | supported; exfoliated; n<5 | 97 |
| graphene | ~2430 | Theory: BTE, $3^{rd}$-order IFCs | $K(graphene) \geq K(carbon\ nanotube)$ | 81 |
| graphene | 1000 - 8000 | Theory: BTE+RTA $\gamma_{LA}, \gamma_{TA}$ | strong size dependence | 84 |
| graphene | 2000-8000 | Theory: BTE+RTA, $\gamma_s(q)$ | strong edge, width and grunaisen parameter dependence | 66 |
| graphene | ~ 4000 | Theory: ballistic | strong width dependence | 50 |
| graphene | 500 - 1100 | Theory: molecular dynamic, optimized Tersoff | T ~ 435 K, calculation domain 4.4 x 4.3 x 1.6 nm³: periodic boundary condition | 99 |
| graphene | ~ 2900 | Theory: MD simulation | strong dependence on the vacancy concentration | 105 |
| graphene | 1500 - 3500 | Theory: BTE, $3^{rd}$-order IFCs | strong size dependence | 107 |
| FLG | 1000 - 3500 | Theory: BTE, $3^{rd}$-order IFCs | N = 1 – 5, strong size dependence | 45 |
| FLG | 2000-3300 | Theory: BTE, $3^{rd}$-order IFCs | n = 1 - 4 | 46 |
| FLG | 580 - 880 | Theory: MD simulation | N = 1 – 5, strong dependence on the Van-der Vaals bond strength | 102 |
| GNR | 1000 - 7000 | Theory: molecular dynamics, Tersoff | strong ribbon width and edge dependence | 49 |
| GNR | ~ 5500 | Theory: BTE + RTA | GNR with width of 5 μm; strong dependence on the edge roughness | 71 |